\documentclass[11pt]{article}
 \usepackage{color}
\usepackage{amsmath}
\usepackage{amssymb}
\usepackage{amsthm}
\usepackage{epstopdf}
\usepackage{graphicx}
\usepackage{mathtools}
\usepackage[usenames,dvipsnames, table]{xcolor}
\usepackage[hang]{caption}
\usepackage{hyperref}

\usepackage{oldgerm}  

\def\bee{\begin{enumerate}}\def\eee{\end{enumerate}}
\def\bei{\begin{itemize}}\def\eei{\end{itemize}}
\newcommand{\nco}{\newcommand}
\def\CC{{\cal C}}

\nco{\red}{\color{red}}
\nco{\blue}{\color{blue}}
\nco{\cyan}{\color{cyan}}
\nco{\brown}{\color{Magenta}}

\def\Red#1{\red #1\normalcolor}
\nco{\magenta}{\color{magenta}}

\nco{\violet}{\color{violet}}
\nco{\redend}{\normalcolor}
\def\inv#1{\frac{1}{#1}}
\def\cy{\#{\rm cycles}}

\def\tr{{\rm tr}\,}
\def\ie{{\it i.e. }}
\def\ommit#1{{}}
\def\({\left(}
\def\){\right)}
\def\oh{\frac{1}{2}}
\def\ie{{\it i.e.,\/}\ }
\def\ie{{\rm i.e.,\/}\ }

\def\be{\begin{equation}}\def\ee{\end{equation}}
\def\bea{\begin{eqnarray}}\def\eea{\end{eqnarray}}

\def\SU{\mathrm{SU}}
\def\Ud{U^\dagger}
\def\Jd{K}
\def\Z{\Bbb{Z}}
 \def\Cat{{\rm Cat}}
 \def\z{z}\def\d{d}

\nco{\rnc}{\renewcommand}
\rnc{\title}[1]{{\Large\bf\mbox{}\\\medskip#1\bigskip\medskip\\}}
\rnc{\author}[1]{{\large #1\smallskip\\}}
\nco{\address}[1]{{\em #1\medskip\\}}

\def\bbeta{\tilde\kappa}

\begin{document}

\begin{titlepage}


\begin{center}
\title{Revisiting $\SU(N)$ integrals }
\medskip
\author{Jean-Bernard Zuber}
\address{
 Sorbonne Universit\'es, UPMC Univ Paris 06, UMR 7589, LPTHE, F-75005, 
Paris, France\\
\& CNRS, UMR 7589, LPTHE, F-75005, Paris, France
 }
\bigskip\medskip

\begin{abstract}
\noindent In this note, I revisit integrals over $\SU(N)$ of the form $ \int DU\, U_{i_1j_1}\cdots U_{i_pj_p}\Ud_{k_1l_1}\cdots \Ud_{k_nl_n}$. While the case $p=n$ is well known, it seems that explicit expressions for $p=n+N$ had not 
appeared
in the literature. Similarities and differences, in particular in the large $N$ limit, between the two cases are discussed.
\end{abstract}
\end{center}

\vspace*{70mm}
\end{titlepage}

\section{Introduction and results}
In this  note, we consider the $\SU(N)$ integrals
\be\label{intnp} \int DU\, U_{i_1j_1}\cdots U_{i_pj_p}\Ud_{k_1l_1}\cdots \Ud_{k_nl_n}\,,\ee
with $DU$ the normalized Haar measure, 
or their generating functions
\be\label{genfnZnp}Z_{p,n}(J,\Jd)=\int DU\,  (\tr (\Jd U))^p  (\tr(J \Ud))^n \,,\ee
where $J$ and $K$ are arbitrary $N\times N$ matrices. 

Such integrals, mainly over the group U$(N)$, 
have been the object of numerous publications in the past, in the context of 
lattice gauge theories \cite{Weingarten, Creutz, others, DrouffeZ}, or in the large $N$ limit \cite{BrezinGross,OBZ}, 
or for their connections with combinatorics \cite{Collins, PZJ}. Integrals over $\SU(N)$ 
seem to have received less attention, see, however, \cite{Creutz, Carlsson}. Recent work by Rossi and Veneziano \cite{RV} 
has prompted  this new investigation.

\medskip
{Let us first recall the physical motivations for studying the $\SU(N)$ integrals (\ref{intnp}) or
(\ref{genfnZnp}).  
Such integrals appear in the context of  
lattice calculations of baryon spectrum.
 Indeed, consider a  SU$(N)$ lattice gauge theory, with link variables denoted $U_\ell\in \SU(N)$,
 Wilson action ${\mathcal{S}}=\beta \sum_{\rm plaquettes} \tr U_P$, and lattice averages
$\langle \cdot \rangle:= \int \prod_\ell DU_{\ell} (\cdot) e^{\mathcal{S}}/ \int \prod_{\ell} DU_\ell  e^{\mathcal{S}}$.
Following  \cite{RV}, introduce the baryonic Wilson loop or ``book observable" 
 $$ \langle {\mathcal{B}}\rangle  := \langle  \epsilon_{i_1\cdots i_N} \epsilon_{j_1\cdots j_N}
 \prod_{a=1}^N U^{(a)}_{i_aj_a}  \rangle$$
 where the ordered products $U^{(a)}=\prod_{\ell \in \CC^{(a)}} U_\ell$, $a=1,\cdots N$, stand for $N$  (static) quark lines 
 joining two points A and B, a distance $r$ apart, see Fig. 1. ${\mathcal{B}}$ represents a baryon made of $N$ quarks, 
 created at A and annihilated at B. 
To lowest order in a small $\beta$  (strong coupling) expansion, $\langle {\mathcal{B}}\rangle=\beta^{\rm \mathcal{A}}
\epsilon_{i_1\cdots i_N} \epsilon_{j_1\cdots j_N}
 \int DV \prod_{a=1}^N V_{i_aj_a} 
$, with $ \mathcal{A}$ the total area of the $N$ ``sheets" of the book, and the last $V$-integration is carried out 
on the ``junction" of these sheets and is given by $Z_{N,0}$.
 Higher order corrections in $\beta$ may involve some other integrals $Z_{n+N,n}$. 
}
\begin{figure}[h]
\begin{center}\includegraphics[width=6cm]{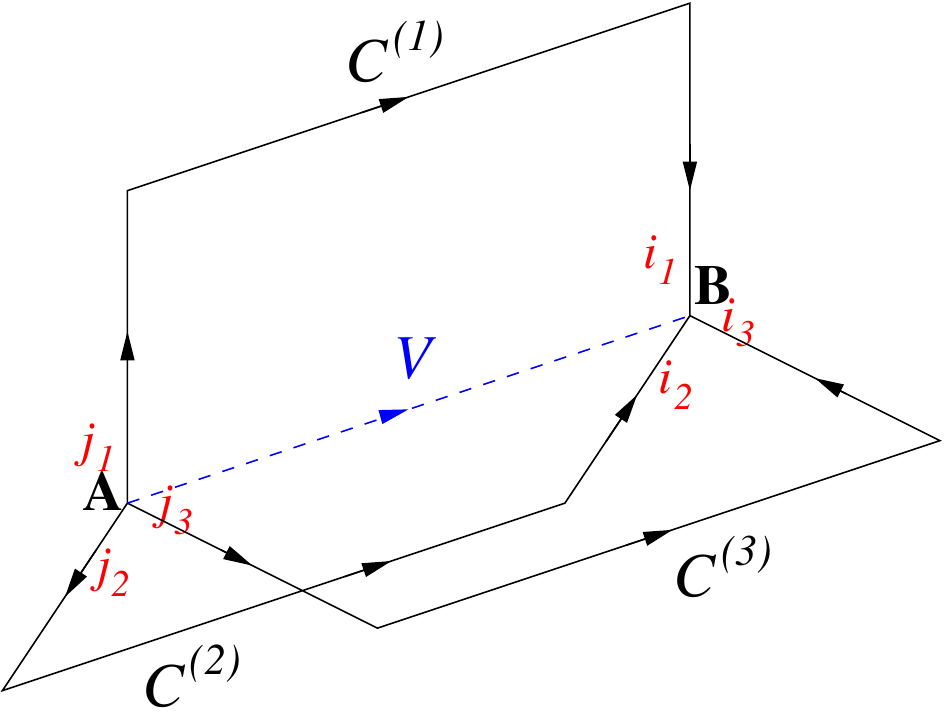}
\caption{A book observable for $N=3$.}\end{center}
\end{figure}

\medskip

 By $\Z_N$ invariance, it is clear that the above integrals  vanish if
\be\label{non-van-cond} p-n\ne 0\ \mod N\,. \ee
 From a representation theoretic point of view, the number of independent terms in (\ref{genfnZnp}),
 (\ie of independent tensors with the right symmetries in (\ref{intnp})),  is given by the number of invariants 
 in $(N)^{\otimes p} \otimes (\bar N)^{\otimes n} $, where $(N)$ and $(\bar N)$ denote the fundamental $N$-dimensional
 representation of $\SU(N)$ and its complex conjugate. 

\medskip
 $Z_{n,n}(J,\Jd)$ is a well known function of (traces of powers of) $J\Jd$, (``Weingarten's function" \cite{Weingarten}),
 at least for $n<N$,   and one may 
 collect all $Z_{n,n}$'s into
\be\label{defZW} Z_W(J,\Jd)=\sum_{n\ge 0} \frac{\kappa^{2n}}{(n!)^2}  Z_{n,n}(J,K)
\,.\ee
For the convenience of the reader, a certain number of known results on these integrals and their generating function 
are recalled in the Appendix. This case $n=p$ in eqs. (\ref{intnp}), (\ref{genfnZnp}) will be referred to as ``the ordinary case".

\medskip
We now turn to the determination of the $Z_{n+N,n}$.  First, for $n=0$,
$\int DU\, U_{i_1j_1}\cdots U_{i_Nj_N}$ is given by the only invariant in $\otimes^N (N)$, 
namely by the totally antisymmetric tensor product, (\ie the determinant of $U$, equal to 1 in $\SU(N)$).
Hence 
$$\int DU\, U_{i_1j_1}\cdots U_{i_Nj_N} =A  \epsilon_{i_1\cdots i_N}\epsilon_{j_1\cdots j_N}$$
with a constant $A$ determined by contraction with $\epsilon_{i_1\cdots i_N}$ and use
of $\epsilon_{i_1\cdots i_N} U_{i_1j_1}\cdots U_{i_Nj_N}= \epsilon_{j_1\cdots j_N}\det U$,
hence $\int DU\,\det U=\int DU=1= A\epsilon_{j_1\cdots j_N}\epsilon_{j_1\cdots j_N}=A N!$,
and $A=1/N!$. Thus
\be\label{int0} \int DU\, U_{i_1j_1}\cdots U_{i_Nj_N} =\inv{N!} \epsilon_{i_1\cdots i_N}\epsilon_{j_1\cdots j_N}\,.\ee
Equivalently, 
\be Z_{N,0}(J,\Jd)=\det \Jd\,. \ee

 More generally,  for $n<N$, $Z_{N+n,n}(J,K) $ is the product of $\det K$ by a polynomial  of $M:=J\Jd$, invariant under 
 $M\mapsto VMV^\dagger$, for any $V\in \SU(N)$, and homogeneous of degree $n$ in $M$, hence a polynomial that 
 may be expanded on traces of powers of $M$
\be\label{ZNnn} Z_{N+n,n} = \int DU\, (\tr (U\Jd))^{N+n} \,(\tr (\Ud J))^n= \det \Jd \, \sum_{\alpha\,\vdash\, n} \d_\alpha t_\alpha\, . \ee
with a sum 
over partitions of $n$ denoted $\alpha=[1^{\alpha_1} , 2^{\alpha_2},\cdots, p^{\alpha_p}]$, $\sum_q q \,\alpha_q=n$,
and with the notations
\be t_\alpha:= 
\prod_q  t_q^{\alpha_q} \,, \qquad t_q:= \tr (J\Jd)^q\,.\ee
The coefficients $\d_\alpha$ are determined through recursion formulae resulting from a contraction of (\ref{intnp}) with
a Kronecker delta: 
\be\label{diff-eqn}
\delta_{jk}\frac{\partial^2}{\partial J_{lk} \partial \Jd_{ji}} Z_{N+n,n}=(N+n) n \,\delta_{il}\,Z_{N+n-1,n-1}\,.\ee
\\
The first coefficients are  readily determined
\bea\nonumber
n=1 & \d_{[1]}=1 \\ \nonumber
n=2 & \d_{[2]}=-\frac{1}{N}\qquad \d_{[1^2]}=\frac{N+1}{N} \\ \label{coeff123}
n=3 & \d_{[3]}= \frac{4}{N(N-1)} \qquad \d_{[1,2]}=-\frac{3(N+1)}{N(N-1)} \qquad 
\d_{[1^3]}= \frac{(N+1)^2-2}{N(N-1)}   \\ 
\nonumber n=4 &  \d_{[4]}=-\frac{30}{N(N-1)(N-2)}
\qquad \d_{[1,3]}=\frac{8(2(N + 1)^2 - 3)}{(N+1)N(N-1)(N-2)} 
\qquad \d_{[2^2]}=
\frac{3 ((N+1)^2+6)}{(N+1)N(N-1)(N-2)} \qquad\\
\nonumber   & \d_{[1^2,2]}=-\frac{6((N+ 1)^2 - 4)}{N(N-1)(N-2)} \qquad \d_{[1^4]}=\frac{(N+ 1)^4 - 8 (N + 1)^2 + 6}{(N+1)N(N-1)(N-2)} \qquad  \\
\nonumber {\rm etc.}
\eea

These coefficients are in fact simply related to the analogous coefficients $\z_\alpha$ in the expansion of the ordinary generating
function, see Appendix, eq. (\ref{Zalphadef}).
If $\z_\alpha$, $\alpha \vdash n$,  is written in the form
\be\label{Zgener} \z_\alpha =\frac{P_\alpha(N)}{N^2 (N^2-1)\cdots (N^2-(n-1)^2)}\ee
with $P_\alpha(N)$ a polynomial of $N$, 
then \be\label{Dgener}\d_\alpha =\frac{P_\alpha(N+1)}{(N+1)\cdots (N-(n-2))}\,. \ee
This follows from the comparison between the two systems of recursion formulae, see below sect. \ref{recurr}, and
may be verified on the first coefficients (\ref{coeff123}) and (\ref{Z1234}).

\medskip
Note  that the coefficients $\d_\alpha$ decrease  more slowly than the $\z_\alpha$, for fixed $n$,
 as $N$ grows. This behavior has consequences 
on the large $N$ limit of the generating function $Z_D$ defined by
\be\label{defZD} Z_D=\sum_{n\ge 0} \frac{\kappa^n}{n!} Z_{N+n,n}=\int DU\, e^{\kappa\, \tr U\Jd \,\tr \Ud J }\, (\tr U\Jd)^N 
=: \tilde Z_D \det \Jd\,.\ee
Note also the different summations in (\ref{defZW}) and (\ref{defZD}): here $\kappa$ is a homogeneity parameter for the eigenvalues
of $M=J\Jd$ while in (\ref{defZW}), it is $\kappa^2$ that plays that role. In both cases, however, we take $\kappa$ of order $N$
and set $\kappa=N\bbeta$. 
Then while in the ordinary case, a non trivial limit is obtained by taking 
the traces $t_n$ also 
 of order $N$, resulting in an exponentation $Z_W=\exp N^2 W_W$, here one finds that the traces $t_n$ have to be 
 kept of order 1 and then $\tilde Z_D=\exp N W_D$, with $W_D$ a non trivial function of the $t_n$'s.

Indeed one finds for the first terms
 \be \lim_{N\to \infty} \inv{N} \log \frac{Z_D}{\det \Jd}=
\tilde\kappa t_1 +\frac{\tilde\kappa^2}{2} (t_1^2-t_2) +\frac{\tilde\kappa^3}{3}(t_1^3-3t_1t_2+2t_3) +\frac{\tilde\kappa^4}{4}(t_1^4-6t_2t_1^2 +2t_2^2 +8t_3t_1-5t_4)+\cdots
 \ee
which is just the beginning of a simple formula in the large $N$ limit: 
\be\label{simple-form} W_D:=\lim_{N\to \infty} \inv{N} \log \frac{Z_D}{\det \Jd}=
\sum_{n\ge1}\tilde\kappa^n \, \sum_{\alpha\,\vdash\, n} (-1)^{n-\sum \alpha_q} \frac{(n-1)!}{(n-\sum \alpha_q+1)!}
\prod_p   \frac{(\Cat(p-1)\, t_p)^{\alpha_p}}{\alpha_p!} 
\ee
in terms of  the Catalan numbers $\Cat(m)=\frac{(2m)!}{m!(m+1)!}$. 
This will be proved below in sect. \ref{expon}.

\section{The recursion relations}
\label{recurr}
\def\wa{\widehat \alpha_{(q)}}
\def\wwa{\widehat{\widehat \alpha}_{(q)}}
\def\ta{\widetilde\alpha_{(q,r)}}

Let us return to (\ref{diff-eqn}) and look at the action of the differential operator $\frac{\partial}{\partial \Jd_{ji}} $ on a typical term
$t_\alpha \det \Jd$ of (\ref{ZNnn}).   
Making use of the identities
\bea \frac{\partial}{\partial \Jd_{ji}} \det[\Jd] &=& ({\rm Cof} \Jd)_{ji}\\
 ({\rm Cof} \Jd)_{ji}  (\Jd X)_{jl} &=& (\det \Jd)  X_{il} 
 \eea
 one may write for each term in the expansion of $Z_{N+n,n}$
 \bea \nonumber 
 &&\frac{\partial^2}{\partial \Jd_{ji} \partial J_{lj}} (t_\alpha \det \Jd)
 = \frac{\partial}{\partial \Jd_{ji}}  \sum_{q=1}^p q \alpha_q   (\Jd(J\Jd)^{q-1})_{jl} 
  \,t_{\wa} \det \Jd
 \\ \label{recurs}
 &=& \sum_{q=1}^p q \alpha_q \Big( (N+1)  (J\Jd)^{q-1}_{il} +\sum_{s=1}^{q-1} t_s  (J\Jd)^{q-s-1}_{il} \Big) t_{\wa} \det \Jd
\\ &+& \nonumber   \sum_{q=1}^p q^2 \alpha_q( \alpha_q-1)  (J\Jd)^{2q-1}_{il} t_{\wwa} \det \Jd 
+2  \sum_{1\le q < r\le p} qr\alpha_q\alpha_r  (J\Jd)^{q+r-1}_{il}\, t_{\ta} \det \Jd 
\eea
with $\wa:=(1^{\alpha_1}\cdots q^{\alpha_q-1}\cdots p^{\alpha_p})$,  
$\wwa:=(1^{\alpha_1}\cdots q^{\alpha_q-2}\cdots p^{\alpha_p})$, 
and $\ta:=(1^{\alpha_1}\cdots q^{\alpha_q-1}\cdots r^{\alpha_r-1}\cdots p^{\alpha_p})$.

In this way, (\ref{diff-eqn}) yields 
 an (overdetermined) system of relations between the coefficients $\d_\alpha$ at ranks $n$ and $n-1$, namely
\bea\label{recurD}  &&\sum_{\alpha\,\vdash\, n} \d_{\alpha}\Big\{ 
 \sum_{q=1}^p q \alpha_q \Big(  \Red{(N+1)}  (J\Jd)^{q-1}_{il} +\sum_{s=1}^{q-1} t_s  (J\Jd)^{q-s-1}_{il} \Big) t_{\wa} \det(\Jd)
 \Big. \\ \Big.
 &+& \nonumber   \sum_{q=1}^p q^2 \alpha_q( \alpha_q-1)  (J\Jd)^{2q-1}_{il} t_{\wwa} \det(\Jd)
+2  \sum_{1\le q < r\le p} qr\alpha_q\alpha_r  (J\Jd)^{q+r-1}_{il}\, t_{\ta} \det(\Jd) \Big\}\\ \nonumber
&&\qquad = n \Red{(N+n)} \delta_{il} \sum_{\alpha'\,\vdash\, n-1} \d_{\alpha'} t_{\alpha'} \det(\Jd)\, .
\eea

Compare these equations with those satisfied by the coefficients $\z_\alpha$ in the similar expansion of
 $Z_{n,n}(J,\Jd)$ in the ordinary case, see (\ref{recurZ}). Their structure is the same, except for changes
 in the  terms of (\ref{recurD}) marked in red. 
In the  linear system on the $\d_\alpha$, the parameter $N$ in the lhs of (\ref{recurZ}) has been changed  into $N+1$ while the 
right hand side is multiplied by $(N+n)$. As a result  
the solutions of the $\d_\alpha$ linear system are  obtained from those of the $\z_\alpha$ one by
$$  {\rm for\ a\  given\ } n \qquad \d_\alpha =(N+n)(N+n-1)\cdots (N+1) \Big( \z_\alpha |_{N\to N+1}\Big)\,.$$
If $\z_\alpha$ is written as in (\ref{Zgener}), it follows that $\d_\alpha$ has the form (\ref{Dgener}), qed. 
\\
In particular, this relation implies that $\z_\alpha$ and $\d_\alpha$ have the same overall sign, namely
 $(-1)^{\#{\rm cycles}(\alpha)+n}$.

 
\section{ Exponentiation and large $N$ limit}
\label{expon}

   The differential equation (\ref{diff-eqn}) carries over to the generating function $Z_D$ of (\ref{defZD}) in the form
   \be \delta_{jk}\frac{\partial^2}{\partial J_{lk} \partial \Jd_{ji}} Z_D=\delta_{il}\, \Big((N+1) \kappa + \kappa^2 \frac{\partial}{\partial \kappa}\Big)Z_D\,.\ee
   We write $Z_D=\det K\, \tilde Z_D$, in which $\tilde Z_D$ is a function of $M:=J\,K$ invariant under $M\to V M V^\dagger$
   for any $V\in \SU(N)$, and we rewrite the differential equation as
   \be \frac{\partial^2 \tilde Z_D}{\partial M_{lk}\partial M_{ji}}  M_{jk} +(N+1) \frac{\partial \tilde Z_D}{\partial M_{li}}=
   \delta_{il}\, \Big((N+1) \kappa + \kappa^2 \frac{\partial}{\partial \kappa}\Big)\tilde Z_D\,.\ee
This may be reexpressed as a differential equation wrt  the eigenvalues $\lambda_i$ of $M$ (generically $M$ is diagonalizable).
This is a standard procedure \cite{BrezinGross} with the result that for any $i$, $1\le i\le N$
\be \lambda_i \frac{\partial^2 \tilde Z_D}{\partial \lambda_i^2}+ (N+1) \frac{\partial \tilde Z_D}{\partial \lambda_i}
+\sum_{j\ne i} \lambda_j \frac{\frac{\partial \tilde Z_D}{\partial \lambda_i}-\frac{\partial \tilde Z_D}{\partial \lambda_j}}{\lambda_i-\lambda_j}= \Big((N+1) \kappa + \kappa^2 \frac{\partial}{\partial \kappa}\Big)\tilde Z_D\,. \ee
   Finally we write $\tilde Z_D=\exp N W_D$, which results in 
\be \lambda_i \Big( N \frac{\partial^2  W_D}{\partial \lambda_i^2}
+ N^2 \left(\frac{\partial  W_D}{\partial \lambda_i}\right)^2\Big)
+ (N+1)N \frac{\partial W_D}{\partial \lambda_i}
+N\sum_{j\ne i} \lambda_j \frac{\frac{\partial W_D}{\partial \lambda_i}-\frac{\partial W_D}{\partial \lambda_j}}{\lambda_i-\lambda_j}= (N+1) \kappa + N\kappa^2 \frac{\partial W_D}{\partial \kappa}\,. \ee
In the large $N$ limit, we rescale $\kappa=N\bbeta$, keeping all $t_n=\sum_i\lambda_i^n$ of order 1, and after dropping 
the subdominant terms, we get for $w_i:= \frac{\partial W_D}{\partial \lambda_i}$
the equation
\be \lambda_i w_i^2 + w_i = \bbeta +\bbeta^2 \frac{\partial W_D}{\partial \bbeta}\,. \ee
(Note that this is in contrast with the ``ordinary case" where the term $\sum_{j\ne i} \cdots$ contributes 
in the large $N$ limit \cite{BrezinGross}.)
 Now $\bbeta$ is just an homogeneity variable of the   $\lambda$'s, and we may thus 
 substitute $\bbeta \frac{\partial W_D}{\partial \bbeta}= \sum_j \lambda_j \frac{\partial W_D}{\partial \lambda_j}=\sum_j \lambda_j w_j$.
 The equation finally reduces to a system of {\it algebraic} equations for the $w_i$'s
 \be\label{} \lambda_i w_i^2 + w_i = \bbeta (1+\sum_j \lambda_j w_j)\,. \ee
Assuming $w:=\sum_i \lambda_i w_i$ known, one finds
\bea\label{wi-eqn}  
\lambda_i w_i^2 + w_i = \bbeta (1+w)\\ \nonumber
w_i= \frac{-1+\sqrt{1+4\lambda_i \bbeta(1+w)}}{2\lambda_i}\eea
and $1+w$ is thus the root of 
\be \label{w-eqn}1+w=1-\frac{N}{2} +\sum_i \frac{\sqrt{1+4\lambda_i \bbeta(1+w)}}{2}\,.\ee
 
 Equation (\ref{wi-eqn}) should be compared with that of the generating function of Catalan numbers $\Cat(m)=\frac{(2m)!}{m!(m+1)!}$, namely 
 $C(t)=\sum_{n\ge 0} \Cat(n) t^n$, $tC^2(t) - C(t)+1=0$. We find that $w_i=\bbeta (1+w) C(-\bbeta(1+w)\lambda_i)$,
 hence 
 \be w_i= \bbeta (1+w)\sum_{n\ge 0} \Cat(n) (-\bbeta(1+w)\lambda_i)^n\ee
 and
 \be\label{w-eqn2} w=\sum_i \lambda_i w_i= -\sum_{n\ge 1} \Cat(n-1) (-\bbeta(1+w))^n t_n\,.\ee

 Let $f_0=1$, $f_n:= (-1)^{n-1}  \Cat(n-1) t_n$ for $n\ge 1$, $y:= \tilde \kappa (1+w)$, then (\ref{w-eqn2}) reads
 $y =\tilde \kappa\sum_{n\ge 0} f_n y^n$, whose solution is given by Lagrange formula
 \bea\nonumber y=\tilde \kappa(1+w) &=&\sum_{n\ge 0}  \frac{\tilde\kappa^{n+1}}{(n+1)!}   
\Big( \frac{d}{dz}\Big)^n\Big( \sum_{m\ge 0} f_m z^m \Big)^{n+1}\Big|_{z=0}\\
\nonumber &=&\tilde\kappa + \sum_{n\ge 1 } \frac{\tilde\kappa^{n+1}}{(n+1)!} \sum_{\alpha\vdash n}\frac{(n+1)!}{(n+1-\sum_q\alpha_q)! \prod_q \alpha_q!} n! f_\alpha\\
\nonumber w=\sum_{n\ge 1}  m_n \tilde\kappa^n &=& \sum_{n\ge 1} \tilde\kappa^n \sum_{\alpha\vdash n}\frac{n!}{(n+1-\sum_q\alpha_q)! \prod_q \alpha_q!}  f_\alpha \eea
where the multinomial coefficient  appears naturally in the expansion of the $n+1$-th power,   and the $n!$ 
results from the  $n$-th derivative of $z^n$. 
Upon integration we get
$$W_D= \sum_{n\ge 1} \tilde\kappa^n \sum_{\alpha\vdash n}(-1)^{n-\sum\alpha_q}\frac{(n-1)!}{(n+1-\sum_q\alpha_q)! \prod_q \alpha_q!}  
\prod_p (\Cat(p-1) t_p)^{\alpha_p}$$ 
which establishes (\ref{simple-form}).

 \ommit{Now from the above mentioned study of the relationship between ordinary moments $m_n$  and planar cumulants
 $f_n$,
 it is well known that the functional inverse of the generating function 
 $P(z)=\inv{z} \Big( 1+\sum_{n\ge 1} f_n z^n  \Big)$ is 
 $G(u)=\inv{u}(1+\sum_{n\ge 1} m_n u^{-n})$ with
 \be m_n =\sum_{\alpha\vdash n}\frac{n!}{(n+1-\sum_q \alpha_q)!} \prod_p \frac{f_p^{\alpha_p}}{\alpha_p!}\,.\ee
  Here we take 
 $f_n=-\Cat(n-1) t_n \bbeta^n$ and we thus conclude that the solution $(1+w)$ of (\ref{w-eqn}) 
\bea 1+w &=& 1+ \sum_{n\ge1} (-1)^{n-1} \Cat(n-1) \bbeta^n t_n (1+w)^n\\
 &=& -(1+w) P(-(1+w))\,,\eea 
 \ie $P(-(1+w))=-1$, is \be -(1+w)=G(-1)\ee
 or
 $w=\sum_{n\ge1} (-1)^n m_n\,. $
 This is just equivalent to the expression (\ref{simple-form}) for $W_D$. 
 \hskip 3cm QED}
 
 Note the similarity of this calculation with the relation between  ordinary moments $m_n$  and 
 non crossing (or free) cumulants
 $f_n$ of a given distribution.
 The combinatorial or diagrammatical  interpretation of  (\ref{simple-form}) remains to be found. 
 
\ommit{ \bigskip
 \magenta{Questions}\\
 Question 1: why is it natural to have $\log (Z/\det \Jd)$ of order $N$?\\
 Question 2: what are these symmetric functions (of the eigenvalues of $J\Jd$) that appear in the limit of $W_D$ ?
Start like the elementary symm fns, but differ already at $n=4$ \dots \\
Question 3: what's the meaning of all $t$'s of order 1 rather than $N$ ? $M=J\Jd$ of finite rank ?\\
Question 4: combinatorial or diagrammatical  interpretation of  (\ref{simple-form}) ?
\normalcolor}

\section*{Acknowledgements} 
It is a pleasure to thank Gabriele Veneziano for inspiring discussions.


 \section*{Appendix}
  \setcounter{equation}{0} 
 \renewcommand{\theequation}{A.\arabic{equation}}
 In this Appendix, we recall a certain number of results on 
 the Weingarten's functions
 $Z_{n,n}(J,\Jd)$ and their generating function $Z_W(J,\Jd)$ of (\ref{defZW}).
The $Z_{n,n}$ are well known functions of (traces of powers of) $J\Jd$, at least for $n<N$, 
see also \cite{Creutz, others,DrouffeZ}.  For $n\ge N$, the traces are no longer independent, or equivalently 
the tensors $\prod \delta_{i \pi(l)} \delta_{j \rho(k)}$, $\pi, \rho\in S_n$, are no longer independent, and we have to
deal with a ``pseudo-inverse" of the Gram matrix, see \cite{Collins, PZJ}.

 \bigskip 
Consider the integrals on 
 U$(N)$ (or SU$(N)$, this is irrelevant here, assuming $n<N$)
\cite{Weingarten, DrouffeZ, Collins}
\bea 
\int DU\, U_{i_1j_1}\cdots U_{i_nj_n} \Ud_{k_1\ell_1} \cdots\Ud_{k_n\ell_n}&=&
\sum_{\tau,\sigma\in S_n} C([\sigma])
\prod_{a=1}^n \delta_{i_a\ell_{\tau(a)}} \delta_{j_a k_{\tau\sigma(a)}}\\
&=&\nonumber
\sum_{\tau,\sigma\in S_n}
\sum_{Y {\rm Young\ diagr.}\atop |Y|=n}
{(\chi^{(\lambda)}(1))^2 \chi^{(\lambda)}([\sigma])\over n!^2 s_\lambda(I)}
\prod_{q=1}^p \delta_{i_q\ell_{\tau(q)}} \delta_{j_q k_{\tau\sigma(q)}}
\eea 
where $\chi^{(\lambda)}([\sigma])$ is the character of the symmetric group 
$S_n$ associated with the Young  diagram $Y$, (a function of the 
class  $[\sigma]$ of $\sigma$);  thus $\chi^{(\lambda)}(1)$ is the
dimension of that representation; $s_\lambda(X)$ is the character 
of the linear group GL($N$) associated with Young diagram $Y$, 
that is a Schur function when expressed in terms of the eigenvalues of $X$; 
$s_\lambda(I)$  is thus the dimension of that representation. Finally the 
coefficient $C([\sigma])$ will be determined below.

Alternatively, in terms of generating functions with sources $J$ and $\Jd$
\be\label{weing} Z_{n,n}(J,\Jd)=\int DU (\tr \Jd U)^n (\tr J \Ud)^n = \sum_{\alpha\vdash n} 
n! |\alpha| \,C([\alpha])\, t_\alpha\\  \ee
where 
$|\alpha|$ is the cardinal of class $[\alpha]$ in $S_n$, thus $|\alpha|=\frac{n!}{\prod_p p^{\alpha_p} \alpha_p !}$.
In (\ref{weing}), the factor $n!$ comes from the sum over $\tau$ and the factor $|\alpha|$ from that over 
the elements $\sigma\in [\alpha]$. 
Thus 
\be\label{Zalphadef} 
 Z_W(J,\Jd):= \int_{U(N)} DU \exp[\kappa \tr(\Jd U+J\Ud)] =\sum_{n=0}^\infty \frac{\kappa^{2n}}{n!} \sum_{\alpha\vdash n} \z_\alpha  t_\alpha\\  \ee
with $\z_\alpha=|\alpha|\, C([\alpha])$. 
 
 By the same argument as in sect. 2, the $\z_\alpha$ satisfy the linear system of recursion relations
\bea\label{recurZ}  &&\sum_{\alpha\vdash n} \z_{\alpha}\Big\{ 
 \sum_{q=1}^p q \alpha_q \Big(  N  (J\Jd)^{q-1}_{il} +\sum_{s=1}^{q-1} t_s  (J\Jd)^{q-s-1}_{il} \Big) t_{\wa} 
 \Big. \\ \nonumber \Big.
 &&\qquad +    \sum_{q=1}^p q^2 \alpha_q( \alpha_q-1)  (J\Jd)^{2q-1}_{il} t_{\wwa} 
+2  \sum_{1\le q < r\le p} qr\alpha_q\alpha_r  (J\Jd)^{q+r-1}_{il}\, t_{\ta}  \Big\}\\ \nonumber
&&\qquad = n  \delta_{il} \sum_{\alpha'\,\vdash\, n-1} \z_{\alpha'} t_{\alpha'}\, .
\eea

Explicitly, the first coefficients  $\z_\alpha=|\alpha| C([\alpha])$ read 
\bea\nonumber
n=1 & {\z_{[1]}=\frac{1}{N}}  
\\ \nonumber
n=2 & {\z_{[{2}]}=-\frac{1}{(N^2 - 1) N}\ ,\quad \z_{[{1^2}]}=\frac{1}{(N^2 - 1) }}
\\ \label{Z1234}
n=3 & 
{\z_{[3]}=\frac{4}{(N^2 -  4) (N^2 - 1) N }
\ ,\quad \z_{[{1,2}]}=-\frac{3 }{(N^2 -  4) (N^2 - 1) }
\ ,\quad
\z_{[1^3]} 
=\frac{N^2 -   2}{(N^2 -  4) (N^2 - 1) N }
} 
 \\  \nonumber n=4 & 
 {\z_{[{4}]}=-\frac{30}{(N^2- 9) (N^2 - 4) (N^2 - 1) N}
\ ,\quad
\z_{[{1,3}]}=\frac{8(2N^2 - 3)}{(N^2- 9) (N^2 - 4) (N^2 - 1) N^2}
} 
\\
\nonumber   & 
{\z_{[{2^2}]} 
=\frac{3(N^2 +  6)}{(N^2- 9) (N^2 - 4) (N^2 - 1) N^2} 
}\,,\ \
\z_{[{1^2,2}]}
=-\frac{6(N^2-4)}{(N^2- 9) (N^2 - 4)(N^2 - 1) N}
\,,  
\\ \nonumber &
{\z_{[{1^4}]}
=\frac{N^4 - 8 N^2 +   6}{(N^2- 9) (N^2 - 4) (N^2 - 1) N^2}}\\
\nonumber {\rm etc.}
\eea
The overall sign of $\z_\alpha$ is $(-1)^{\cy(\alpha)-n}$ and its large $N$ behavior $|\z_\alpha|\sim {N^{-2n+\cy(\alpha)}}$.

\bigskip
To study the 
large $N$ limit, we take $\kappa=N\tilde \kappa$  and 
 $t_p=N \tau_p$, with $\tilde \kappa$ and $\tau_p$ of order 1.
Then Br\'ezin and Gross   \cite{BrezinGross} have shown that 
\be  W_W(J\Jd):= \lim_{N\to \infty} \inv{N^2} \log Z_W(J,  \Jd;N \tilde\kappa)\ee
exists and satisfies the coupled equations
\bea \label{BGsystem}
W_W &=&\frac{2}{N} \sum_i (\tilde\kappa^2 \lambda_i+c)^\oh -\inv{2N^2} \sum_{i,j}\log[(\tilde\kappa^2 \lambda_i+c)^\oh
+(\tilde\kappa^2 \lambda_j+c)^\oh]-c-\frac{3}{4}\\
{\rm with }\quad c&=&
\begin{cases} \inv{N} \sum_i (\tilde\kappa^2 \lambda_i+c)^{-\oh} & {\rm for}\  \inv{N} \sum_i(\tilde\kappa^2 \lambda_i)^{-\oh} \ge 2
\qquad {\rm (``strong\ coupling")}\\
0 &  {\rm for}\  \inv{N} \sum_i(\tilde\kappa^2 \lambda_i)^{-\oh} \le 2  \qquad {\rm (``weak\ coupling")} \end{cases} 
\eea
The solution has two determinations, in a strong coupling and in a weak coupling phase.
Here we are concerned with the strong coupling regime in which we may expand
\be W_W=\sum_{n\ge 1} \tilde\kappa^{2n} \sum_{\alpha\,\vdash\, n} w_\alpha \tau_\alpha\ee
with the first terms 
\be W_W=\tilde\kappa^2 \tau_1 + \frac{\tilde\kappa^4}{2} (\tau_1^2 -\tau_2) +\frac{2\tilde\kappa^6}{3}(2 \tau_1^3 - 3 \tau_1 \tau_2 + \tau_3)
+\frac{\tilde\kappa^8}{4} (24 \tau_1^4 - 48 \tau_1^2 \tau_2 + 9 \tau_2^2 + 
  20 \tau_1 \tau_3 - 5 \tau_4)+\cdots
 \ee 
 and the general term  given in \cite{OBZ}
\be\label{OBZres}
w_\alpha= (-1)^n 
 \frac{(2n-3+\sum_q\alpha_q)!}{(2n)!} \prod_{q} \left(-\frac{(2q)!}{(q!)^2}\right)^{\alpha_q} \inv{\alpha_q !} 
\,.\ee

 \end{document}